\documentclass[
final            
  ,numberedheadings 
  ]
  {aipproc}

\usepackage{amsmath}
\usepackage{colortbl}
\usepackage{anysize}
\usepackage{array}
\usepackage{latexsym, amscd, amsthm}
\usepackage{pdfsync}
\usepackage{epsf}
\usepackage{amssymb}

\layoutstyle{8x11single}

\DeclareSymbolFont{extraup}{U}{zavm}{m}{n}
\DeclareMathSymbol{\varheart}{\mathalpha}{extraup}{86}
\DeclareMathSymbol{\vardiamond}{\mathalpha}{extraup}{87}
\newcommand{\bdiamond}{{\scriptstyle \vardiamond}}
\newcommand{\cinf}{{{\cal \cC}^\infty(M,\R)}}

\newcommand{\tr}{\mathrm{tr}}
\newcommand{\End}{\mathrm{End}}

\newcommand{\Mat}{\mathrm{Mat}}
\newcommand{\ev}{\mathrm{ev}}
\newcommand{\eqdef}{\stackrel{{\rm def.}}{=}}

\DeclareFontFamily{U}{rsf}{}
\DeclareFontShape{U}{rsf}{m}{n}{<5> <6> rsfs5 <7> <8> <9> rsfs7 <10-> rsfs10}{}
\DeclareMathAlphabet\Scr{U}{rsf}{m}{n}

\newcommand{\KA}{K\"{a}hler-Atiyah~}
\newcommand{\Aord}{{A=\mathrm{ordered}}}

\def\C{{\mathbb{C}}}
\def\R{{\mathbb{R}}}

\def\H{{\mathbb{H}}}
\def\Z{{\mathbb{Z}}}
\def\S{{\mathbb{S}}}
\def\rk{{\rm{rk}}}

\def\vol{\mathrm{vol}}
\def\AdS{\mathrm{AdS}}

\newcommand{\id}{\mathrm{id}}
\newcommand{\be}{\begin{equation}}
\newcommand{\ee}{\end{equation}}
\newcommand{\beqa}{\begin{eqnarray}}
\newcommand{\eeqa}{\end{eqnarray}}
\newcommand{\nn}{\nonumber}

\newcommand{\bE}{\mathbf{E}}

\def\cR{{\mathcal R}}
\def\cP{\mathcal{P}}
\def\cC{{\mathcal C}}
\def\cB{\Scr B}

\def\Cl{\mathrm{Cl}}
\def\odd{\mathrm{odd}}

\def\cC{\mathcal{C}}
\newcommand{\pqarray}{\begin{array}{c} p-q\\ {\rm mod}~8\end{array}}
\newcommand{\twopartdef}[4]
{
	\left\{
		\begin{array}{ll}
			#1 & \mbox{if } #2 \\
			#3 & \mbox{if } #4
		\end{array}
	\right.
}

\begin{document}

\title{A unified approach to Fierz identities}


\author{E. M. Babalic}{
  address={IFIN-HH, Department of Theoretical Physics, 077125  Magurele, Romania}
}

\author{I. A. Coman}{
  address={IFIN-HH, Department of Theoretical Physics, 077125  Magurele, Romania}
}

\author{C. I. Lazaroiu}{
  address={IFIN-HH, Department of Theoretical Physics, 077125  Magurele, Romania}
  ,altaddress={IBS Center for Geometry and Physics, and POSTECH, Dept. of Math., Pohang, Gyeongbuk 790-784, Korea} 
}

\begin{abstract}
We summarize a unified and computationally efficient treatment of
Fierz identities for form-valued pinor bilinears in various dimensions
and signatures, using concepts and techniques borrowed from a certain
approach to spinors known as ``geometric algebra''. Our formulation 
displays the real, complex and quaternionic structures in a
conceptually clear manner, which is moreover amenable to
implementation in various symbolic computation systems.
\end{abstract}

\maketitle


\section{Introduction}

Computations involving Fierz identities in curved backgrounds for
various dimensions and signatures are a cumbersome ingredient of
supergravity and string theory and their applications.  This problem
can be aleviated by using geometric algebra techniques, which afford a
unified treatment of Fierz identities for form-valued pinor bilinears.
Our study \cite{gf:2013,ga1:2013} connects previous work 
(\cite{Okubo1:1995,Okubo2:1991,Rand:1992,AC1:1997,AC2:2005}) 
with techniques and ideas belonging to the theory of \KA
bundles and modules over such, otherwise known as ``geometric algebra''.

\

\noindent{{\bf Notations and conventions.}} Let $(M,g)$ denote any smooth, 
connected and {\em oriented} pseudo-Riemannian manifold of dimension
$d=p+q$, where $p$ and $q$ are the numbers of positive and negative
eigenvalues of the metric tensor $g$.  We further assume that $M$ is
paracompact, so that we have partitions of unity subordinate to any
open cover. The space of $\R$-valued smooth inhomogeneous and globally-defined differential 
forms on $M$ is a $\Z$-graded $\cinf$-module denoted $\Omega(M)\eqdef \Gamma(M,\wedge T^\ast M)$, 
with fixed rank components $\Omega^k (M)=\Gamma(M,\wedge^k T^\ast M)$ for $k=0,\ldots ,d$.
The (real) volume form of $(M,g)$ is denoted by $\nu=\vol_M\in \Omega^d(M)$ and satisfies 
the following properties:
\beqa
\label{NuSquared}
\nu\diamond
\nu &=&(-1)^{q+\left[\frac{d}{2}\right]}1_M=\twopartdef{(-1)^{\frac{p-q}{2}}
  1_M~,~}{d={\rm even}}{(-1)^{\frac{p-q-1}{2}}1_M~,}{d=\odd~}~,\nn\\
  \nu \diamond \omega &=& \pi^{d-1}(\omega)\diamond \nu ~~~,~~~\forall \omega\in \Omega(M)~,\nn
\eeqa
with respect to the \emph{geometric product} $\diamond$ (see \cite{ga1:2013}). Hence $\nu$ is central 
in the \KA algebra $(\Omega(M),\diamond)$ when $d$ is odd and twisted central 
(i.e., $\nu \diamond \omega=\pi(\omega)\diamond \nu$, where $\pi$ is 
the \emph{grading} or \emph{main} automorphism) when $d$ is even. In Table 
\ref{table:AlgClassif}, at the intersection of each row and column of the first sub-table, 
we indicate the values of $p-q~({\rm mod}~8)$ for which the volume form $\nu$ has the
corresponding properties.  Some general aspects of the geometric algebra formalism which we use 
here can also be found in \cite{Tim1:2013} and are discussed in detail in \cite{gf:2013, ga1:2013}.

A real \emph{pinor bundle} on $(M,g)$ is defined as an $\R$-vector
bundle endowed with a morphism of bundles of algebras $\gamma:(\wedge
T^\ast M,\diamond)\rightarrow(\End(S),\circ)$ which turns $S$ into a
bundle of modules over the the \KA bundle of $(M,g)$, which is the
exterior bundle $\wedge T^*M$ endowed with the geometric product
$\diamond$.  A real \emph{pin bundle} is a pinor bundle for which $S$
is a bundle of simple modules over the \KA bundle.  A real
\emph{spin(or) bundle} is a bundle $S$ of (simple) modules 
over the {\em even rank sub-bundle} $\wedge ^\ev T^\ast M$ of
the \KA bundle.  

\

\noindent{\bf Spin projectors and spin bundles.} 
Giving a direct sum bundle decomposition $S=S_+\oplus S_-$  amounts to
giving a {\em product structure} on $S$, a nontrivial globally-defined bundle
 endomorphism $\cR\in \Gamma(M,\End(S))\setminus \{-\id_S,\id_S\}$ satisfying:
\be
\label{Rprops1}
\cR^2=\id_S~~~~.\nn 
\ee
A product structure $\cR$ is called a \emph{spin endomorphism} if it also satisfies the condition:
\be
[\cR,\gamma(\omega)]_{-,\circ}=0~~,~~\forall \omega\in \Omega^\ev(M)~~.\nn
\ee
A spin endomorphism exists only when $p-q\equiv_8 0,4,6,7$. When $S$ is a pin bundle, the 
restriction $\gamma_\ev\eqdef \gamma|_{\wedge^\ev T^\ast M}:
(\wedge^\ev T^\ast M,\diamond) \rightarrow (\End(S),\circ)$ is
fiberwise reducible iff. $S$ admits a spin endomorphism $\cR$, in
which case we define the {\em spin projectors} determined by $\cR$ to
be the globally-defined endomorphisms $\cP^\cR_\pm\eqdef \frac{1}{2}(\id_S \pm \cR)$,
which are complementary idempotents in $\Gamma(M,\End(S))$. Thus the eigen-subbundles $S^\pm\eqdef \cP^\cR_\pm (S)$
corresponding to the eigenvalues $\pm1$ of $\cR$ are complementary in $S$, i.e. $S=S^+\oplus S^-$, and 
$\cR$ determines a nontrivial direct sum decomposition $\gamma_\ev=\gamma^+\oplus \gamma^-$.

\

\noindent{\bf The effective domain of definition of $\gamma$.} 
Let $\wedge^\pm T^\ast M$ denote the bundle of twisted (anti-)selfdual forms \cite{ga1:2013}. Its space 
$\Omega^\pm(M)\eqdef \Gamma(M,\wedge^\pm T^\ast M)$ of smooth global sections is 
the $\cinf$-module consisting of those forms $\omega\in \Omega(M)$ which satisfy the condition ~$\omega\diamond \nu=\pm \omega$. ~Defining: 
\beqa
\wedge^{\gamma} T^\ast M&\eqdef& \twopartdef{\wedge T^\ast M~~,~~}{~\gamma~~\mathrm{is~fiberwise~injective~(simple~case)}~,}{\wedge^{\epsilon_\gamma} T^\ast M~,~~}
{~\gamma~~\mathrm{is~not~fiberwise~injective~(non-simple~case)}~,}~~\nn\\
\wedge^{-\gamma} T^\ast M&\eqdef& \twopartdef{0~~,~~}{~\gamma~~\mathrm{is~fiberwise~injective~(simple~case)}~,}
{\wedge^{-\epsilon_\gamma} T^\ast M~~,~~}{~\gamma~~\mathrm{is~not~fiberwise~injective~(non-simple~case)}~,}~~\nn
\eeqa
one finds that $\gamma$ restricts to zero on $\wedge^{-\gamma} T^\ast M$ and to a monomorphism of vector bundles on 
$\wedge^\gamma T^\ast M$. 

\section{Schur algebras and representation types}

\noindent{\bf Definition.} Let $S$ be a pin bundle of $(M,g)$ and $x$ be any point of $M$. The {\em Schur
algebra} of $\gamma_x$ is the unital subalgebra $\Sigma_{\gamma,x}$ of $(\End(S_x),\circ)$
defined through:
\be
\Sigma_{\gamma, x}\eqdef \{T_x\in \End (S_x)~|~ [T_x,\gamma_x(\omega_x)]_{-.\circ}=0~,
~\forall \omega_x\in \wedge T^\ast_x M \}~~.\nn
\ee
The subset~ $\Sigma_\gamma=\{(x,T_x)~|~x\in M~,~ T_x\in \Sigma_{\gamma,x}\}=\sqcup_{x\in M}\Sigma_{\gamma,x}$~ 
is a sub-bundle of unital algebras of the bundle of algebras
$(\End(S),\circ)$, which we shall call the {\em Schur bundle} of
$\gamma$. The isomorphism type of the fiber $(\Sigma_{\gamma,
  x},\circ_x)$ is independent of $x$ and is denoted by $\S$, being
called the {\em Schur algebra} of $\gamma$.  Real pin bundles $S$ are
of three types: {\em normal}, {\em almost complex} or {\em
  quaternionic}, depending on whether their Schur algebra $\S$ is
isomorphic with $\R$, $\C$ or $\H$. We summarize some of their
properties in Table \ref{table:types_extended}.  Since $\gamma$ is
fiberwise irreducible in the cases of interest below, its Schur
algebra $\S$ depends only on $p-q~({\rm mod}~8)$. In Tables
\ref{table:AlgClassif}, we indicate in parentheses the corresponding
Schur algebras. Note that the real Clifford algebra $\Cl(p,q)$ is
non-simple iff. $p-q\equiv_8 1,5$, which we indicate in tables
through the blue shading.

\vspace{1mm}

\begin{table}[h]
\begin{tabular}{c c}
 \begin{tabular}{|c |c |c|}
\hline
~~ & $\nu\diamond\nu=+1$ & $\nu\diamond\nu=-1$ \\
\hline
$\nu$ is central & \cellcolor{cyan}$\mathbf{1(\R),5(\H)}$ & $\mathbf{3(\C)}, {\mathbf{7(\C)}}$ \\
\hline
$\nu$ is not central & ${\mathbf{0(\R),4(\H)}}$ & $\mathbf{2(\R)}, {\mathbf{6(\H)}}$ \\
\hline
\end{tabular}
&
\begin{tabular}{|c|c|c|}
\hline
~ & ~injective~ & ~non-injective~ \\
\hline
surjective     & $ \mathbf{0(\R)}, \mathbf{2(\R)}$ & \cellcolor{cyan} $\mathbf{1(\R)}$ \\
\hline
non-surjective & $\mathbf{3(\C)}, \mathbf{7(\C), 4(\H), 6(\H)}$ & \cellcolor{cyan} $\mathbf{5(\H)}$ \\
\hline
\end{tabular}
\caption{Properties of $\nu$ according to $p-q~(\mathrm{mod}~8)$ and
fiberwise character of real pin representations $\gamma$.}
\label{table:AlgClassif}
\end{tabular}
\end{table}

\begin{table}[ht]
\begin{tabular}{|c|c|c|c|c|c|c|c|}
\hline
$\S$ & $\begin{array}{c} p-q\\ {\rm mod}~8 \end{array}$ 
& $\begin{array}{c}\wedge T^\ast_x M\\\approx \Cl(p,q)\end{array}$ & $\Delta$ & $N$ & 
$\begin{array}{c} \mathrm{Number~of}\\ \mathrm{choices}~ \mathrm{for}~\gamma \end{array}$   
& $\gamma_x(\wedge T^\ast_x M)$ & $\begin{array}{c} \mathrm{Fiberwise}\\ \mathrm{injectivity}~ \mathrm{of}~\gamma\end{array}$\\
\hline\hline
$\R$ & $\mathbin{\textcolor{red}{\mathbf{0}}}, \mathbf{2}$ & $\Mat(\Delta,\R)$ & $2^{[\frac{d}{2}]}=2^{\frac{d}{2}}$ & $2^{[\frac{d}{2}]}$ & $1$ & $\Mat(\Delta,\R)$   & injective \\
\hline
$\H$ & $\mathbin{\textcolor{red}{\mathbf{4, 6}}}$ & $\Mat(\Delta,\H)$ & $2^{[\frac{d}{2}]-1}=2^{\frac{d}{2}-1}$ & $2^{[\frac{d}{2}]+1}$ & $1$ & $\Mat(\Delta,\H)$   & injective\\
\hline
$\C$ & $\mathbf{3}, \mathbin{\textcolor{red}{\mathbf{7}}}$ & $\Mat(\Delta,\C)$ & $2^{[\frac{d}{2}]}=2^{\frac{d-1}{2}}$ &$2^{[\frac{d}{2}]+1}$& 1 & $\Mat(\Delta,\C)$    & injective\\
\hline
$\H$ & $\mathbin{\textcolor{blue}{\mathbf{5}}}$ & $\Mat(\Delta,\H)^{\oplus 2}$ & $2^{[\frac{d}{2}]-1}=2^{\frac{d-3}{2}}$ &$2^{[\frac{d}{2}]+1}$&$2$ $(\epsilon_\gamma=\pm 1)$&$\Mat(\Delta,\H)$& 
\cellcolor{cyan}non-injective \\
\hline
$\R$ & $\mathbin{\textcolor{blue}{\mathbf{1}}}$ & $\Mat(\Delta,\R)^{\oplus 2}$ & $2^{[\frac{d}{2}]}=2^{\frac{d-1}{2}}$ & $2^{[\frac{d}{2}]}$& $2$ $(\epsilon_\gamma=\pm 1)$ & $\Mat(\Delta,\R)$ & 
\cellcolor{cyan} non-injective \\
\hline 
\end{tabular}
\caption{Summary of pin bundle types. 
$N\eqdef \rk_\R S$ is the real rank of $S$ while $\Delta\eqdef
\rk_{\Sigma_\gamma}S$ is the Schur rank of $S$. The non-simple cases are indicated through 
the blue shading of the corresponding table cells. The red color indicates those cases
for which a spin endomorphism can be defined.}
\label{table:types_extended}
\end{table}

\vspace{1mm}

Well-known facts from the representation theory of Clifford algebras imply the following:

\noindent 1. $\gamma$ is \emph{fiberwise injective} iff. $\Cl(p,q)$ is simple as an associative
$\R$-algebra, i.e. iff. $p-q\not \equiv_8 1,5$ (called \emph{simple case}). 

\noindent 2. When $\gamma$ is fiberwise \emph{non-injective} (i.e. when $p-q\equiv_8
1,5$, the so-called {\em non-simple case}), we have $\gamma(\nu)=\epsilon_\gamma \id_S$,
where the sign factor $\epsilon_\gamma\in \{-1,1\}$ is called the {\em
signature} of $\gamma$. The two choices for $\epsilon_\gamma$ lead to
two inequivalent choices for $\gamma$. 
The fiberwise injectivity and surjectivity of $\gamma$ are summarized in 
the second sub-table of Table \ref{table:AlgClassif}.

\section{Geometric Fierz identities for real pinors}

\paragraph{{\bf Admissible bilinear pairings}}
A non-degenerate bilinear pairing $\cB$ on $S$ is called {\em
admissible} \cite{AC1:1997,AC2:2005} if:

\begin{enumerate}

\item $\cB$ is either symmetric or skew-symmetric, i.e. $\cB(\xi,\xi')=\sigma_\cB \cB(\xi', \xi),~~\forall \xi,\xi' \in \Gamma(M,S)$,~
with {\em symmetry} factor $\sigma_\cB \in\{-1,+1\}$);

\item For any $\omega\in \Omega(M)$, we have:
\beqa
\label{Btype}
&& \gamma(\omega)^t=\gamma(\tau_\cB(\omega))~\Longleftrightarrow ~
\cB(\gamma(\omega)\xi,\xi')= \cB(\xi,\gamma(\tau_\cB(\omega)) \xi')~~,~~\forall \xi,\xi' \in \Gamma(M,S)~~,\\
\label{tauBdef}
\!\!\!\! \mathrm{where} &&\tau_\cB\eqdef \tau\circ \pi^{\frac{1-\epsilon_\cB}{2}} = \twopartdef{\tau~,~}{\epsilon_\cB=+1}{\tau\circ \pi~,~}{\epsilon_\cB=-1}~
~~(\mathrm{with}~\tau(\omega^{(k)})=(-1)^{\frac{k(k-1)}{2}},~\forall \omega^{(k)}\in\Omega^k(M))
\eeqa
is the {\em $\cB$-modified reversion} and the sign factor $\epsilon_\cB \in
\{-1,1\}$ is called the {\em type} of $\cB$;
 
\item If $p-q\equiv_8 0,4,6,7$ (thus $S=S^+ \oplus S^-$ where $S^\pm\subset S$ are real spin bundles), then $S^+$ and $S^-$ are either $\cB$-orthogonal 
to each other or $\cB$-isotropic. The {\em isotropy} of $\cB$ is the sign factor $\iota_\cB\in \{-1,1\}$ defined through:
\be
\label{Bisotropy}
 \iota_\cB \eqdef \twopartdef{+1~,~}{\cB(S^+,S^-)=0}{-1~,~}{\cB(S^\pm,S^\pm)=0}~~.\nn
\ee
When $p-q\not \equiv 0,4,6,7$, the isotropy $\iota_\cB$ is undefined. 
\end{enumerate}
The number and properties of independent admissible bilinear pairings (studied in detail in \cite{AC1:1997,AC2:2005}) depend on 
$p$ and $q$.

\

\noindent{\bf Local expressions.} Let $e^m$ be a pseudo-orthonormal local coframe of $(M,g)$
defined above an open subset $U\subset M$. Then property \eqref{Btype}
amounts to $(\gamma^m)^t=\epsilon_\cB\gamma^m~\Longleftrightarrow~  \cB(\gamma^m\xi,\xi')=\epsilon_\cB \cB(\xi,\gamma^m\xi'),~~\forall m=1\ldots d$,
which in turn implies:
\beqa
\label{gammaAtranspose}
&&(\gamma^A)^t=\epsilon_\cB^{|A|} (-1)^{\frac{|A|(|A|-1)}{2}}\gamma^{A}~~,~~~\forall A=(m_1,...,m_k),~\mathrm{with}~1\leq m_1<\ldots < m_k\leq d~~,\\
\label{typeA}
&&(\gamma^A)^{-t}=\epsilon_\cB^{|A|}\gamma_A~~\Longleftrightarrow ~~
\cB((\gamma^A)^{-1}\xi,\xi')=\epsilon_\cB^{|A|} \cB(\xi,\gamma_A\xi')~~,
~~\forall \xi,\xi'\in \Gamma(M,S)~~.
\eeqa
If $(\varepsilon_i)_{i=1\ldots N}$ is an arbitrary local frame of $S$
defined above $U$ (with dual local frame $(\varepsilon^i)_{i=1\ldots
  N}$ of $S^\ast$), then:
\be
\label{Tij}
T|_U\varepsilon_i=\sum_{j=1}^N T_{ji}\varepsilon^j=\sum_{j=1}^N \varepsilon^j(T\varepsilon_i)\epsilon_j ~~,~~\forall T\in \Gamma(M,\End(S))~~,
~~\mathrm{where}~T_{ij}\eqdef \varepsilon^i(T|_U\varepsilon_j)\in \cC^\infty(U,\R)~.
\ee 

\

\noindent{\bf Preparations.} Given an admissible fiberwise bilinear pairing $\cB$ on $S$, we define
endomorphisms $E_{\xi,\xi'}$ of $S$ through
$E_{\xi,\xi'}(\xi'')\stackrel{\rm{def}}{=}\cB(\xi'',\xi')\xi$ for any
$\xi,\xi'\in \Gamma(M,S)$ (see \cite{ga1:2013}).  Then the following
identities are satisfied:
\beqa
\label{mainid}
 &&E_{\xi_1,\xi_2}\circ E_{\xi_3,\xi_4}=\cB(\xi_3,\xi_2)E_{\xi_1,\xi_4}~~,~~\forall \xi_1,\xi_2,\xi_3,\xi_4 \in \Gamma(M,S)~~,\\
 \label{traceid}
 &&\tr(T\circ E_{\xi,\xi'})=\cB(T\xi,\xi')~~,~~\forall \xi,\xi' \in \Gamma(M,S)~~.
\eeqa

\subsection{Normal case}

\noindent This occurs when $\S\approx \R$, i.e. for $p-q\equiv_8 0,1,2$, in
which case $N=\Delta=2^{\left[\frac{d}{2}\right]}$. It is characterized
by two admissible bilinear pairings $\cB_0,~\cB_1$, which one can take
to be related through $\cB_1=\cB_0\circ(\id\otimes\gamma(\nu))$ and
whose properties are given in \cite{gf:2013,AC1:1997,AC2:2005}. These two pairings
are independent when $p-q\equiv_8 0,2$ (the simple normal cases) and
proportional to each other when $p-q\equiv_8 1$ (the non-simple normal
case).  We summarize some properties of the subcases of the normal case 
in Table \ref{table:NormalCase}.  Here and below, we
use the abbreviations M=Majorana, MW=Majorana-Weyl, SM=symplectic
Majorana, SMW=symplectic Majorana-Weyl, DM=double Majorana for the
(sometimes conflicted) physics terminology. The green shading
indicates those cases for which a spin endomorphism can be defined. We
have $\gamma(\nu)=\gamma^{(d+1)}=\gamma^1\circ\ldots\circ\gamma^d$
in any local positively-oriented pseudo-orthonormal coframe of $(M,g)$.

\begin{table}[ht]
\centering
\begin{tabular}{|c|c|c|c|c|c|c|c|c|c|}
\hline
$\pqarray$ & $\Cl(p,q)$ & $\begin{array}{c}\gamma\\\mathrm{is~injective}\end{array}$  
&$\epsilon_\gamma$ &  $\cR$ (real spinors)& 
$\begin{array}{c}\mathrm{name~of}\\\mathrm{pinors}\end{array}$&$\gamma(\nu)$ & 
$\nu\diamond \nu$ & $\begin{array}{c}\nu\\ \mathrm{is~central}\end{array}$ \\
\hline\hline
$ \mathbin{\textcolor{red}{\mathbf{0}}}$ & simple & Yes & N/A &
\cellcolor{green} $\gamma(\nu)~$ (MW) & M & $\gamma(\nu)$ &$+1$ & No \\
\hline
$ \mathbin{\textcolor{cyan}{\mathbf{1}}}$ & non-simple & No &
\cellcolor{cyan} $\pm 1$ & N/A & M & $\pm 1$ & $+1$ & Yes\\
\hline
$ \mathbf{2}$ & simple & Yes &  N/A & N/A &  M & $\gamma(\nu)$ & $-1$ & No \\
\hline
\end{tabular}
\caption{Summary of subcases of the normal case.}
\label{table:NormalCase}
\end{table} 
Let us start from the local relation \cite{Okubo1:1995}:
 \be
\label{RealOkuboCompleteness}
\sum_\Aord(\gamma^{-1}_A)_{jk}(\gamma_A)_{lm}=\frac{2^d}{N}\delta_{jm}\delta_{lk}~~,\nn
\ee
where $A$ runs over strictly-ordered multi-indices with components from the set $\{1,\ldots,d\}$. 
Multiplying by $T_{kj}$ (see \eqref{Tij}) and summing over $j,k$ gives the 
{\em completeness relation for the normal case}:
\be
\label{realA}
T=_U\frac{N}{2^d}\sum_\Aord \tr(\gamma^{-1}_A \circ T)\gamma_A~~,~~\forall T\in \Gamma(M,\End(S))~~.
\ee
Setting $T=E_{\xi,\xi'}$ in relation \eqref{realA} gives the following expansion upon using \eqref{typeA} and \eqref{traceid}:
\be
\label{RealEExpFinal}
E_{\xi,\xi'}=\frac{N}{2^d}\sum_\Aord \tr(\gamma^{-1}_A \circ E_{\xi,\xi'})\gamma_A=\frac{N}{2^d}\sum_\Aord \epsilon_\cB^{|A|} \cB (\xi,\gamma^A \xi')\gamma_A~~.
\ee
Relation \eqref{RealEExpFinal} implies that the inhomogeneous forms
$\check{E}_{\xi,\xi'}\eqdef \left(\gamma|_{\Omega^\gamma(M)}\right)^{-1}(E_{\xi,\xi'}) \in \Omega^\gamma(M)$
have the following expansion in terms of the \emph{basic admissible pairing} $\cB_0$ :
\be
\check{E}_{\xi,\xi'}=\frac{N}{2^d}\sum_\Aord \epsilon_{\cB_0}^{|A|} 
\cB_0 (\xi,\gamma_A \xi')e^A_\gamma~~,~~\forall \xi,\xi' \in\Gamma(M, S)~~,\nn
\ee
where we used $e^A_\gamma \eqdef \gamma^{-1}(\gamma^A)$. Relation 
\eqref{mainid} implies the \emph{geometric Fierz identities for the normal case}:
\be
\boxed{\check{E}_{\xi_1,\xi_2} \diamond \check{E}_{\xi_3,\xi_4} = 
\cB_0 (\xi_3,\xi_2) \check{E}_{\xi_1,\xi_4}~~,~~\forall \xi_1,\xi_2,\xi_3,\xi_4 \in \Gamma(M,S)}~~.\nn
\ee

\subsubsection{Example: One real pinor in nine Euclidean dimensions}

In this case ($p=9,~q=0$) the pin bundle $S$ is an $\R$-vector bundle
of rank $N=2^{\left[\frac{d}{2}\right]}=16$.  Since $d\equiv_8 1$, and
$p-q\equiv_81$, we are in the normal non-simple case and thus
$\gamma(\nu)=\epsilon_\gamma \id_S$.  Choosing the signature
$\epsilon_\gamma=+ 1$, we realize the subalgebra
$(\Omega^+(M),\diamond)$ of twisted self-dual forms through the
truncated model $(\Omega^<(M),\bdiamond_+)$, where
$\Omega^<(M)=\oplus_{k=0}^4\Omega^k(M)$ and $\bdiamond_+$ is the
reduced geometric product discussed in \cite{ga1:2013}.  Details on the \emph{truncated models} of
the \KA algebra can be found in loc. cit.
Since $\gamma(\nu)=\id_S$ we have only one admissible
pairing $\cB$ on $S$, which has $\sigma_\cB=+1$ and 
$\epsilon_\cB=+1$. We can assume that $\cB$ is
positive-definite and thus is a scalar product on $S$ and we denote
the corresponding norm through $||~||$. 
The isotropy $\iota_\cB$ is not defined.
In the case of one pinor $\xi\in\Gamma(M,S)$ (which we normalize through $||\xi||=1$), we
are interested in pinor bilinears such as
$\check{\bE}^{(k)}\eqdef \frac{1}{k!}\cB(\xi,\gamma_{a_1...a_k}\xi)e^{a_1\ldots a_k}\in\Omega^k(M)~,~\forall a_1,\ldots,a_k\in\overline{1,9}$.
 Using \eqref{gammaAtranspose} and the properties of the bilinear pairing $\cB$, we can construct (up
to twisted Hodge duality on $(M,g)$):
\be
\label{forms9}
\cB(\xi,\xi)=1~~,~~V\eqdef \check{\bE}^{(1)}=\cB(\xi,\gamma_a\xi)e^a ~~
,~~\Phi\eqdef \check{\bE}^{(4)}=\frac{1}{24} \cB(\xi,\gamma_{a_1\ldots a_4}\xi) e^{a_1\ldots a_4}~.\nn
\ee
In this case, the truncated model of the Fierz algebra admits 
a basis consisting of a single element, constructed from the lower truncation of $\check{\bE}$ --- namely  
$\check{E}_<\eqdef \frac{N}{2^d}\check{\bE}_<=\frac{1}{32}\sum_{k=0}^4 \check{\bE}^{(k)}=\frac{1}{32}(1+V+\Phi)$.
The truncated geometric Fierz identity follows easily from \eqref{mainid}, upon using the definition 
of the reduced product $\bdiamond_{+}$ in terms of $\diamond$ (see \cite{ga1:2013}):
\be
\label{Fierz9}
\check{E}_<\bdiamond_{+}\check{E}_<=\frac{1}{2}\check{E}_< ~~\left( ~\Longleftrightarrow~ \check{E}\diamond\check{E}=\check{E}~\right)~~
\Longleftrightarrow~~V\bdiamond_{+ }V+\Phi\bdiamond_{+}\Phi+V\bdiamond_{+}\Phi+\Phi\bdiamond_{+} V=15+14V+14\Phi ~~.
\ee
Solving the system of equations obtained by separating rank components in \eqref{Fierz9} gives, upon using the definition of the 
twisted Hodge star ($\tilde\ast\omega=\omega\diamond\nu$ for any $\omega\in\Omega(M)$),
the following system of conditions on the forms $V$ and $\Phi$:
\be
||V||^2=1 ~~,~~||\Phi||^2=14~~,~~\iota_V\Phi=0~~,~~\tilde\ast(\Phi\wedge\Phi)=14V ~~,
~~\tilde\ast(V\wedge\Phi)=\Phi~~,~~ \Phi\bigtriangleup_2\Phi=-12\Phi\nn~.
\ee

\subsection{Almost complex case}

\noindent This occurs when $\S\approx \C$, which happens for $p-q\equiv_8
3,7$. In this case, $d$ is odd and we have
$N=2\Delta=2^{[\frac{d}{2}]+1}$. There exist two complex structures on the bundle $S$ --- the two {\em
globally-defined} endomorphisms $J\in \Gamma(M,\End(S))$ given by $J=\pm \gamma(\nu)$, which satisfy:
\be
\label{CJconds}
J^2=-\id_S~~~,~~~[J, \gamma(\omega)]_{-,\circ}=0~~,~~\forall \omega\in \Omega(M)~~.\nn
\ee
The results of \cite{Okubo1:1995} imply that there also exists a {\em globally-defined}
endomorphism $D\in \Gamma(M,\End(S))$ which satisfies:
\beqa
&& D\circ \gamma(\omega)=\gamma(\pi(\omega))\circ D~~,~~\forall \omega\in \Omega(M)~~,~~~~~[D,J]_{+,\circ}=0~~,\nn\\
&& D^2=(-1)^{\frac{p-q+1}{4}}\id_S =\twopartdef{-\id_S~,~}{p-q\equiv_8 3~}{+\id_S~,~}{p-q\equiv_8 7~}~~.\nn
\eeqa

\begin{table}[ht]
\centering
\begin{tabular}{|c|c|c|c|c|c|c|c|c|c|}
\hline
$\pqarray$ & $\Cl(p,q)$ & $\begin{array}{c}\gamma\\ \mathrm{is~injective}\end{array}$ & $\epsilon_\gamma$ &  $D^2$ & $\cR$ (real spinors)&
$\begin{array}{c}\mathrm{name~of}\\\mathrm{pinors}\end{array}$&$\gamma(\nu)$ & 
$\nu\diamond \nu$ & $\begin{array}{c}\nu\\ \mathrm{is~central}\end{array}$ \\
\hline\hline
$ \mathbf{3}$ & simple & Yes  & N/A & $-\id_S$ & N/A& M & $\pm J$ & $-1$ & Yes \\
\hline
$ \mathbin{\textcolor{red}{\mathbf{7}}}$ & simple & Yes & N/A & $+\id_S$ &
\cellcolor{green} $D~$ (Majorana)& DM & $\pm J$ & $-1$ & Yes\\
\hline
\end{tabular}
\caption{Summary of subcases of the almost complex case.}
\label{table:ComplexCase}
\end{table} 

\vspace{2mm}

There are four independent choices 
$\cB_0,\cB_1,\cB_2$ and $\cB_3$ for the non-degenerate admissible
pairing, which we can take to be related through $\cB_1=\cB_0\circ (\id_S\otimes J)$~,
~$\cB_2=\cB_0\circ (\id_S\otimes D)$~,~$\cB_3=\cB_0\circ [\id_S\otimes (D\circ J)]$.
Using again the completeness relations of \cite{Okubo1:1995}, a more involved, but similar, derivation to that given in the normal case, 
we obtain (see \cite{gf:2013}) local expansions of the inhomogeneous differential forms which `dequantize' $E_{\xi,\xi'}$  
expressed using the \emph{basic admissible pairing} $\cB_0$. A crucial difference from the normal case is that the bundle morphism 
$\gamma$ is not surjective, having  image equal to $\End_\C(S)=\End_{\Sigma_\gamma}(S)$, where the complex structure on $S$ is defined by $J$. 
To take this into account, we use the fact (see \cite{gf:2013}) that there exits a unique decomposition
$E_{\xi,\xi'}=E^{(0)}_{\xi,\xi'}+D\circ E^{(1)}_{\xi,\xi'}$, with $E^{(0)}_{\xi,\xi'},E^{(1)}_{\xi,\xi'}\in\Gamma(M,\End_\C(S))$,
~ $\forall \xi,\xi'\in\Gamma(M,S)$. Since $\gamma$ is injective, this allows us to define $\check{E}^{(0)}\eqdef \gamma^{-1}(E^{(0)})\in \Omega(M)$ and 
 $\check{E}^{(1)}\eqdef \gamma^{-1}(E^{(1)})\in \Omega(M)$, which have the expansions \cite{gf:2013}:
\beqa 
\check{E}^{(0)}_{\xi,\xi'} &=&_U \frac{\Delta}{2^d}\sum_\Aord (-1)^{|A|}\cB_0(\xi, \gamma^A\xi')e_A~, \nn\\
\check{E}^{(1)}_{\xi,\xi'} &=&_U \frac{\Delta}{2^d}\sum_\Aord (-1)^{\frac{p-q+1}{4}} (-1)^{|A|}\cB_0(\xi, D\circ \gamma^A\xi')e_A ~. \nn
\eeqa
One also finds that \eqref{mainid} implies the \emph{geometric Fierz identities for the almost complex case}:
\be
\boxed{
\begin{split}
& \check{E}^{(0)}_{\xi_1,\xi_2}\diamond \check{E}^{(0)}_{\xi_3,\xi_4} + (-1)^{\frac{p-q+1}{4}} \pi(\check{E}^{(1)}_{\xi_1,\xi_2})\diamond \check{E}^{(1)}_{\xi_3,\xi_4} 
= \cB_0(\xi_3,\xi_2)\check{E}^{(0)}_{\xi_1,\xi_4}~~,\nn\\
& \pi(\check{E}^{(0)}_{\xi_1,\xi_2})\diamond \check{E}^{(1)}_{\xi_3,\xi_4} +  \check{E}^{(1)}_{\xi_1,\xi_2}\diamond \check{E}^{(0)}_{\xi_3,\xi_4} 
= \cB_0(\xi_3,\xi_2)\check{E}^{(1)}_{\xi_1,\xi_4}~~.\nn
\end{split}}
\ee

\subsection{Quaternionic case}

This occurs for $\S\approx \H$, which happens for $p-q\equiv_8
4,5,6$. Then $N=4\Delta=2^{[\frac{d}{2}]+1}$. The Schur algebra is
isomorphic with the $\R$-algebra $\H$ of quaternions, while the Schur
bundle is {\em locally} (over small enough open subsets $U\subset M$)
generated by four linearly-independent elements $J_\alpha\in
\Gamma(U,\End(S))$ ($\alpha=0\ldots 3$) which we can take to
correspond to the quaternion units. Hence $J_0=\id_S$ while
$J_1,J_2,J_3$ satisfy:
\beqa
&& J_i\circ J_j=-\delta_{ij} J_0+\epsilon_{ijk} J_k~~,~~\forall i,j,k =1\ldots 3~\Longrightarrow~ [J_i,J_j]_{+,\circ}=0~~,~~J_i^2=-\id_S~~,\nn\\
&& [J_i,\gamma(\omega)]_{-,\circ}=0~~,~~\forall \omega\in \Omega(U)~~,\nn
\eeqa
where $\epsilon_{ijk}$ is the Levi-Civita symbol. We thus have $\Gamma(U,\End_\H(S))\equiv\{T\in \Gamma(U,\End(S))|[T,J_i]_{-,\circ}=0~,~\forall i=1\ldots 3\}$.

 \vspace{1mm}
 
\begin{table}[ht]
\centering
\begin{tabular}{|c|c|c|c|c|c|c|c|c|c|}
\hline
$\pqarray$ & $\Cl(p,q)$ & $\begin{array}{c}\gamma\\ \mathrm{is~injective}\end{array}$ & $\epsilon_\gamma$ &  $\cR$ (real spinors) &
$\begin{array}{c}\mathrm{name~of}\\\mathrm{pinors}\end{array}$&$\gamma(\nu)$ & 
$\nu \diamond \nu$ & $\begin{array}{c}\nu\\ \mathrm{is~central}\end{array}$ \\
\hline\hline
$ \mathbin{\textcolor{red}{\mathbf{4}}}$ & simple & Yes & N/A &
\cellcolor{green} $\gamma(\nu)~~$ (SMW) & SM & $\gamma(\nu)$ & $+1$ & No \\
\hline
$ \mathbin{\textcolor{cyan}{\mathbf{5}}}$ & non-simple & No & \cellcolor{cyan} $\pm 1$ & N/A & SM & $\pm 1$ &$+1$ & Yes\\
\hline
$ \mathbin{\textcolor{red}{\mathbf{6}}}$ & simple & Yes &  N/A &
\cellcolor{green} $\gamma(\nu)\circ J~~$ (SM) & DM & $\gamma(\nu)$ &$-1$ & No \\
\hline
\end{tabular}
\caption{Summary of subcases of the quaternionic case. $J$ denotes any of the
  complex structures induced on $S$ by the quaternionic structure. }
\label{table:QuaternionicCase}
\end{table} 
In this case, the bundle morphism $\gamma$ has image equal to $\End_\H(S)\eqdef \End_{\Sigma_\gamma}(S)$, where the quaternionic structure of $S$ 
is given locally by $(J_\alpha)_{\alpha=0\ldots 3}$. One can show \cite{gf:2013} that any operator $E_{\xi,\xi'}\in \Gamma(M,\End(S))$ has the unique local decomposition 
$E_{\xi,\xi'}=_U\sum_{\alpha=0}^3 J_\alpha\circ E^{(\alpha)}_{\xi,\xi'}~$ with $E^{(\alpha)}_{\xi,\xi'}=\frac{\Delta}{2^d}\sum_\Aord
\tr(\gamma^{-1}_A\circ J^{-1}_\alpha \circ E_{\xi,\xi'})\gamma_A~\in\Gamma(U,\End_\H(S))$. Therefore, we can define 
$\check{E}^{(\alpha)}_{\xi,\xi'}\eqdef \gamma^{-1}(E^{(\alpha)}_{\xi,\xi'}) \in \Omega(U)$. 
We have eight admissible pairings $\cB^\epsilon_\alpha$ ($\epsilon=\pm 1,~ \alpha=0\ldots 3$), which one can take to be given by
$\cB_k^\epsilon=\cB_0^\epsilon\circ(\id_S\otimes J_k)~,~\forall k=1,\ldots,3$, 
where $\cB^\epsilon_0$ are the so-called {\em basic admissible pairings}. Only four of $\cB^\epsilon_\alpha$  are independent in the quaternionic non-simple case, i.e. 
when $p-q\equiv_8 5$. Fixing a choice for $\epsilon$, we find the local expansions \cite{gf:2013} in terms of the basic admissible pairing:
\beqa
\check{E}^{(0)}_{\xi,\xi'} &=&_U \frac{\Delta}{2^d} \sum_\Aord\epsilon^{|A|}_{\cB_0}\cB_0(\xi, \gamma_A\xi')e^A_\gamma~~,~~\forall \xi,\xi'\in \Gamma(M,S)~~,\nn\\
\check{E}^{(i)}_{\xi,\xi'} &=&_U  \frac{\Delta}{2^d}\sum_\Aord\epsilon^{|A|}_{\cB_0}\cB_0(\xi, (J_i\circ \gamma_A)\xi')e^A_\gamma~~,
~~\forall i=1\ldots 3 ~~\nn
\eeqa
and \eqref{mainid} implies the \emph{geometric Fierz identities for the quaternionic case}:
\be
\boxed{
\begin{split}
  \check{E}^{(0)}_{\xi_1,\xi_2} \diamond  \check{E}^{(0)}_{\xi_3,\xi_4} - \sum_{i=1}^3  \check{E}^{(i)}_{\xi_1,\xi_2} \diamond  \check{E}^{(i)}_{\xi_3,\xi_4} 
&= \cB_0(\xi_3,\xi_2) \check{E}^{(0)}_{\xi_1,\xi_4}~~,\nn\\
  \check{E}^{(0)}_{\xi_1,\xi_2} \diamond  \check{E}^{(i)}_{\xi_3,\xi_4} +  \check{E}^{(i)}_{\xi_1,\xi_2} \diamond  \check{E}^{(0)}_{\xi_3,\xi_4} 
+\sum_{j,k=1}^3 \epsilon_{ijk}  \check{E}^{(j)}_{\xi_1,\xi_2} \diamond  \check{E}^{(k)}_{\xi_3,\xi_4} &= \cB_0(\xi_3,\xi_2) \check{E}^{(i)}_{\xi_1,\xi_4}~~
~(i=1\ldots 3)~.
\end{split}
}
\ee


\begin{theacknowledgments}
This work was supported by the CNCS projects PN-II-RU-TE (contract
number 77/2010) and PN-II-ID-PCE (contract numbers 50/2011 and
121/2011). CIL is supported by the Research Center Program of 
IBS (Institute for Basic Science) in Korea (CA1205-01).
IAC acknowledges her FDP \emph{Open Horizons} scholarship, which financed part of her studies.
\end{theacknowledgments}


\bibliographystyle{aipproc}   

\bibliography{MirelaTimNatbib}


\end{document}